\documentclass[letterpaper,twocolumn,longbibliography,
accepted=2025-04-11]{quantumarticle}
\pdfoutput=1

\usepackage[numbers, sort&compress]{natbib}
\usepackage{mathtools, amssymb}
\usepackage{braket, physics} 

\usepackage{orcidlink}
\graphicspath{{./images/}}

\usepackage{graphicx, placeins, float}
\usepackage[caption=false]{subfig}
\usepackage[export]{adjustbox}

\usepackage{xcolor}
\usepackage{hyperref}
\usepackage[nameinlink,capitalize]{cleveref}
\hypersetup{
	colorlinks = true,
	linkbordercolor = {white},
	linkcolor={purple},
	citecolor={purple},
	urlcolor={blue}
}

\usepackage{algorithm}
\usepackage{algorithmicx}
\usepackage[noend]{algpseudocode}

\crefname{enumi}{Step}{Steps}

\makeatletter
\renewcommand{\ALG@name}{Protocol}
\crefname{algorithm}{Protocol}{Tables}
\makeatother

\DeclareMathOperator{\NOT}{NOT}
\DeclareMathOperator{\REVERSE}{Reverse}

\begin{document}
\title{Interferometric binary phase estimations}
\author{Simone Roncallo\,\orcidlink{0000-0003-3506-9027}}
	\email[]{simone.roncallo01@ateneopv.it}
	\affiliation{Dipartimento di Fisica, Università degli Studi di Pavia, Via Agostino Bassi 6, I-27100, Pavia, Italy}
	\affiliation{INFN Sezione di Pavia, Via Agostino Bassi 6, I-27100, Pavia, Italy}
	
\author{Xi Lu\,\orcidlink{0000-0003-4121-3419}}
	\email[]{luxi@zju.edu.cn}
	\affiliation{School of Mathematical Science, Zhejiang University, Hangzhou, 310027, China}
	
\author{Lorenzo Maccone\,\orcidlink{0000-0002-6729-5312}}
	\email[]{lorenzo.maccone@unipv.it}
	\affiliation{Dipartimento di Fisica, Università degli Studi di Pavia, Via Agostino Bassi 6, I-27100, Pavia, Italy}
	\affiliation{INFN Sezione di Pavia, Via Agostino Bassi 6, I-27100, Pavia, Italy}
	
\begin{abstract}
	We propose an interferometric scheme where each photon returns one bit of the binary expansion of an unknown phase. It sets up a method for estimating the phase value at arbitrary uncertainty. This strategy is global, since it requires no prior information, and it achieves the Heisenberg bound independently of the output statistics. We provide  simulations and a characterization of this architecture.
\end{abstract}
\maketitle

\section{Introduction}
Quantum metrology protocols \citep{art:Giovannetti_Advances,art:Paris,art:Degen} are typically described in terms of an interferometric phase shift. Examples include displacement interferometry \citep{art:Caves}, clock synchronization \citep{art:Komar}, key distribution \citep{art:Bennett,art:Kiyoshi} and imaging \citep{art:Boto,art:Gessner}. Several strategies can estimate a quantum phase, including computational approaches \citep{book:Nielsen,art:Cleve,art:Hassani,art:Cerezo,art:Fitzpatrick}, interferometry \citep{book:Gerry,art:Higgins,art:Smerzi,art:Olivares,art:Sergienko,art:Chuang} and estimation theory \citep{art:Wiebe}. The solution is optimal when it reaches the Heisenberg limit, i.e. when it undergoes the precision scaling set by the quantum bounds \citep{art:Busch,art:Giovannetti_Measurements,art:Gorecki,art:Belliardo}.

In this paper, we introduce an optical protocol for the estimation of an unknown phase value. We design a two-mode interferometric setup, whose response function approximates a square wave. In the ideal case, the mode where the photon exits identifies which among the available intervals the phase value falls into. By iteratively changing the response period, i.e. by controlling the number of phase shifts applied to the photon, we combine all the binary outputs and provide a full phase estimation at arbitrary uncertainty. This step can be performed sequentially, or in parallel without feedforward propagation. Our method can be rephrased as a binary search, which undergoes the Heisenberg scaling in the number of phase shifts experienced by each photon. Such scaling is attained immediately, and not asymptotically, since statistics is not required to complete the search. The protocol requires no prior information, i.e. it is a global estimation method. 

In contrast to learning-based and optimization-based approaches, our strategy does not require training. The shape of the response function is determined by a fixed sequence of parameters. This ensures modularity in controlling the square wave approximation, giving the possibility to extend or shorten the interferometer without further optimization.
\begin{figure}[H]
	\centering
	\includegraphics[width = 0.47 \textwidth]{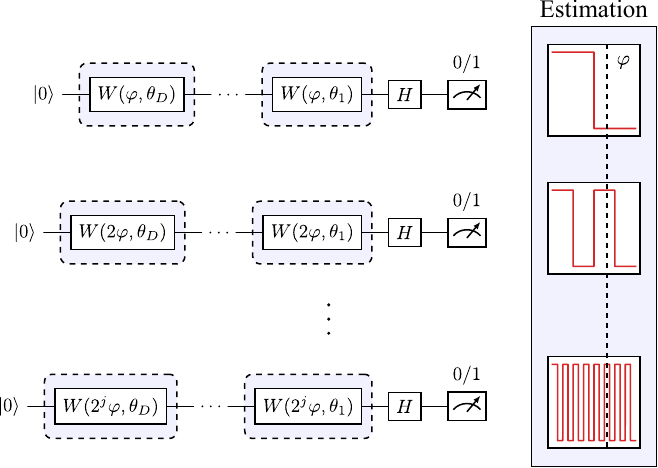}%
	\caption{\label{fig:BinaryEstimation}Binary phase estimation. Each step represents one photon going through the interferometer and providing a single bit of the binary expansion of the unknown phase $\varphi$. At the $j$th step, the output consists of a square-wave response with period $[0,2\pi/2^j)$, with Fourier truncation order determined by the number $D$ of subsequent applications of $W$ (see \cref{fig:Setup}). The true phase is sketched by the dashed vertical line. Combining all the outputs completes the estimation. The depth of the interferometer and the number of steps, i.e. the length of the binary expansion, determines the resource cost of the protocol.}
\end{figure}

\section{Method\label{sec:SecII}}
\begin{figure}
	\centering
	\includegraphics[width = 0.47 \textwidth]{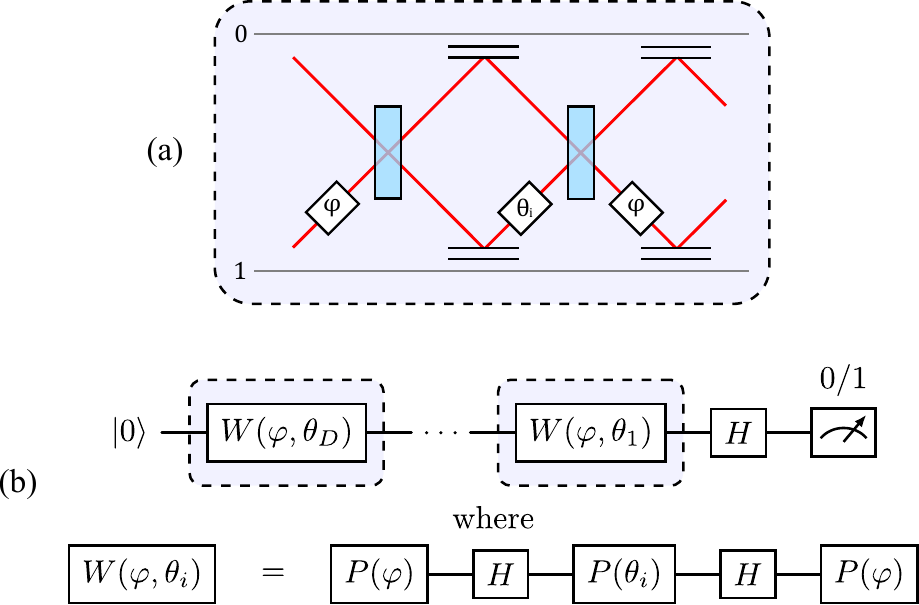}%
	\caption{\label{fig:Setup}Interferometric module. (a) Optical implementation of $W(k_j\varphi, \theta_i)$, for $k_j = 1$. In mode $1$, the photon undergoes three phase transformations, each separated by a beam splitter. Subsequent applications of $W$ (with different $\theta_i$) approximate the truncated Fourier series of the square wave. (b) First iteration ($k_j = 1$) after $D$ subsequent applications of $W$, represented as a single-qubit circuit. The circuit depth determines the Fourier truncation order.}
\end{figure}

We consider the problem of estimating an unknown phase $\varphi \in [0,2\pi)$, encoded by a unitary phase shift $P(\varphi)$. We combine a set of beam splitters and phase transformations into a two-mode interferometer, which takes single-photon states as input (or coherent states, with single-photon component post-selected by a photon-number resolving detector). This is achieved by appending a sequence of unitary blocks $W$ with behaviour controlled by the phase parameters $\{\theta_i\}$. This provides a modular, i.e. extendible, structure. Ideally, the interferometric response function reproduces a square wave. When a photon goes through it, the probability of exiting the first mode yields one bit of the binary expansion of $\varphi$. The square wave period, controlled by the number of subsequent applications of $P(\varphi)$, determines which of the bits is returned. In practice, the interferometric response approximates the truncated Fourier series of the square wave. In this case, the truncation order is determined by the number of subsequent applications of $W$ (with parameters changed at each composition). When $\varphi$ cannot be encoded at different spatial positions, the sequential strategy is substituted by a multiround implementation, in which the photon passes multiple times through the same unknown phase shift \citep{art:Higgins,art:DeCillis}. In this case, the truncation order is fixed by the number of rounds a single photon makes on the same module, with phase parameters coherently changed at each passage. \cref{fig:BinaryEstimation,fig:Setup} illustrate the estimation protocol and the interferometric module design, respectively.

Consider a binary estimation made of $n \in \mathbb{N}^+$ iterations, each associated to an interferometer, whose response function approximates a square wave with period $[0,2\pi/k_j)$. Here, $j \in \{0,\ldots,n-1\}$ and $k_j = 2^{j}$. We adopt the dual-rail encoding, where a photon can propagate in two orthogonal modes, represented by the quantum states $\ket{0}$ and $\ket{1}$. The photon is initially injected in mode $0$. Let $\theta = \{\theta_i\}_{i\in\mathbb{N}}$ be a fixed set of parameters, labelled by $i \in \{1, \ldots, D\}$. As building block of our setup, we define the operator
\begin{equation}
	W(k_j\varphi, \theta_i) = P(k_j\varphi) H P(\theta_i) H P(k_j\varphi) \ , 
	\label{eq:UnitaryModule}
\end{equation}
where $H$ is the Hadamard gate (implemented by a $50\!:\!50$ beam splitter), and $P$ a phase shift transformation. Optically, this operator is equivalent to applying three phase shifts to mode $1$, each interspersed by a beam splitter \citep{art:Lee}. Consider $D$ subsequent applications of $W$ (each with a different parameter $\theta_i$), followed by a final beam splitter (Hadamard). We measure the expectation value of the projector $\Pi_0 = \ket{0}\!\bra{0}$, i.e. the ratio of photons $p_{jD}$ in mode $0$, yielding
\begin{equation}
\begin{gathered}
	p_{jD}(\varphi) = \bra{0} U^{\dagger}(k_j\varphi, \theta) \Pi_0 U(k_j\varphi, \theta) \ket{0} \ , \\
   \text{with} \ U(k_j\varphi, \theta) = H \prod_{i=1}^{D} W(k_j\varphi, \theta_{i}) \ .
\end{gathered}
\label{eq:Probability}
\end{equation}
Since $P(k_j\varphi)\ket{0} = \ket{0}$, it follows that
\begin{gather}
	U(k_j\varphi, \theta) \ket{0} = H P(k_j\varphi) \widetilde{U}(k_j\varphi, \theta)\ket{0} \ , \\
	\widetilde{U}(k_j\varphi, \theta) = L(\theta_1) P(2k_j\varphi) L(\theta_2) \cdots P(2k_j\varphi) L(\theta_D) \ , \label{eq:Decomposition}
\end{gather}
where $L(\theta_i) = H P(\theta_i) H$. By Stone's theorem \citep{art:Stone}, we write $P(2k_j \varphi)$ in terms of its generator $2 k_j \ket{1}\!\bra{1}$, which has spectrum $\{0, k_j \}$. Using the spectral decomposition \citep{art:Schuld}, the response reads
\begin{equation}
	p_{jD}(\varphi) = \sum_{q \in I} c_q(\theta) e^{ik_jq\varphi} \ ,
	\label{eq:FourierTruncated}
\end{equation}
with $I \subseteq \{-2D + 1, -2D + 2, \ldots, 2D-1\}$. This is a truncated Fourier series, with period $\left[0,2\pi/k_j\right)$ and coefficients fixed by $\theta$. Notice that the depth of the interferometer determines the truncation point of the series, i.e. the higher is $D$ the more accurate is the approximation of the square wave. We summarize the whole setup in \cref{fig:Setup}, which represents the first iteration of the binary search estimation. The subsequent iterations are obtained by progressively doubling $k_j$.
\begin{figure*}
	\centering
	\includegraphics[width = 1 \textwidth]{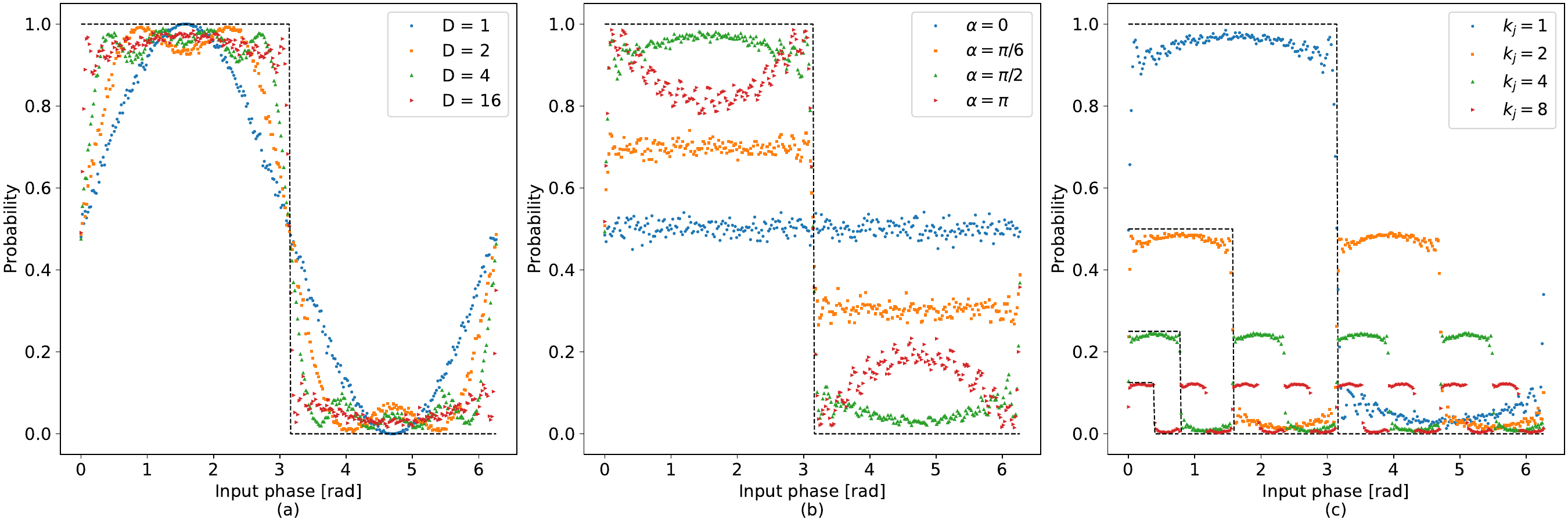}
	\caption{\label{fig:Simulation}Comparison between the square wave and the optical response $p_{jD}$, obtained as the ratio of photons observed in mode $0$. Simulation performed with Qiskit Aer, sampled for $300$ evenly-spaced phases $\varphi \in [0,2\pi)$ and $1000$ shots, for different choices of depth $D$, $\alpha$ and $k_j$. (a) Simulation performed with $\alpha = \pi/2$, $k_j = 1$, and different depths $D$. The response function shows a binary behaviour, with approximation accuracy that increases with the interferometer depth. (b) Simulation with $D = 16$ and $k_j = 1$, plotted for different values of $\alpha$. As $\alpha$ increases, the extrema of $p_{jD}$ approximates $0$ and $1$, up to a critical value, after which the output departs from the expected behaviour. (c) Response function simulated with $D = 16$ and $\alpha = \pi/2$, plotted for different values of $k_j$. As $k_j$ increases, the response period is progressively halved, representing different iterations of the binary search. To enhance visualization, each response is renormalized to $k_j$: all the outputs have true range in $[0,1]$.}
\end{figure*}%

As a generalization of our method, it is possible to train the parameters $\theta$ to approximate an arbitrary response function, with Fourier expressivity controlled by the circuit architecture \citep{book:Petruccione}. Here, we limit to a fixed set of parameters that simulates the square wave. This choice brings several advantages. On the one hand, it bypasses the training procedure, i.e. the additional cost of repeating the protocol for a sufficient amount of optimization epochs. On the other hand, it provides modularity, i.e. it makes possible to extend the interferometer, improving the truncation accuracy without having to reoptimize the former parameters. We evaluate the interferometer response by simulating the circuit of \cref{fig:Setup} when $\theta_i = \alpha/(2i - 1)$, with $\alpha \in \mathbb{R}^+$. The results are reported in \cref{fig:Simulation}. Under this choice of parameters, the output reproduces a square wave. However, this identification holds only approximately. The truncated Fourier series of the square wave admits both negative and greater-than-one values. Such values are strictly forbidden at the interferometer output, which is probabilistic, i.e. $p_{jD}(\varphi) \in [0,1] \ \forall \varphi \in \left[0,2\pi\right)$. This constraint is counterbalanced by $\alpha$, which can be set to saturate the infimum and supremum of $p_{jD}(\varphi)$ to $0$ and $1$, respectively. See \cref{app:Comparison} for a discussion. Alternatively, $\alpha$ can be optimized to mitigate uncertainties or optical imperfections, e.g. using the mean squared error between the ideal and the experimental responses or information metrics on the unknown phase.

We outline the binary estimation of the unknown phase $\varphi$, at arbitrary uncertainty $\varepsilon$. Consider a sequence of square waves $\{f_{j}\}_{j \in \mathbb{N}}$, obtained through the periodic extension of the step function
\begin{equation}
	f_{j}(\varphi) = \begin{cases}
		1 \ \text{for} \ \varphi \in \left[0,\frac{\pi}{k_j}\right) \\[1ex]
		0 \ \text{for} \ \varphi \in \left[\frac{\pi}{k_j},\frac{2\pi}{k_j}\right)
	\end{cases} \ ,
\end{equation}
simulated by $p_{jD}(\varphi)$ at the output of the interferometer. At each iteration, $f_{j}$ partitions the interval $[0,2\pi)$ in $2^{j+1}$ subsets of size $\pi/k_j$, namely
\begin{equation}
	\left\{\left[\tfrac{m\pi}{k_j},\tfrac{(m+1)\pi}{k_j}\right)\right\}_m \ \text{with} \ m \in \{0,1,\ldots,k_{j+1}-1\} \ .
	\label{eq:BinaryIntervalsSequence}
\end{equation}
These subsets are associated to an alternating sequence of $0$s and $1$s, which correspond to the values of $f_{j}$. Combining them identifies which subset includes the true value of the unknown phase. In practice, we do this by measuring the ratio of photons in mode $0$ at the output of the interferometer. The result is rounded to the nearest binary digit, with threshold at $1/2$. The process is repeated, each time doubling the value of $k_j$, which subsequently halves the response period as $f_{j+1}(\varphi) = f_{j}(2\varphi)$. The estimation halts as soon as the partition size becomes double the uncertainty, i.e. $2\varepsilon = \pi/k_{n-1}$ and the number of iterations is $n = \log_2(\pi/\varepsilon)$. If $n \notin \mathbb{N}^+$ we round it down to the first integer, such that $2\varepsilon = \inf_j{\pi/k_j}$. Denoting $\widetilde{m}$ the index of the decision outcome, we estimate the unknown phase $\varphi$ as
\begin{equation}
	\widetilde{\varphi} = \frac{(2\widetilde{m}+1)\pi}{2^n} \pm \frac{\pi}{2^n} \ . 
	\label{eq:BinaryEstimationOutcome}
\end{equation}
The full method is outlined in \cref{alg:Algorithm}. The above procedure can also be parallelized using $n$ interferometers, each set to a different value of $k_j$. Although the parallel strategy brings no advantage in terms of resource counting, as we discuss below, it does not require the feedforward propagation of the sequential one. 
\begin{figure*}
	\centering
	\includegraphics[width = 1 \textwidth]{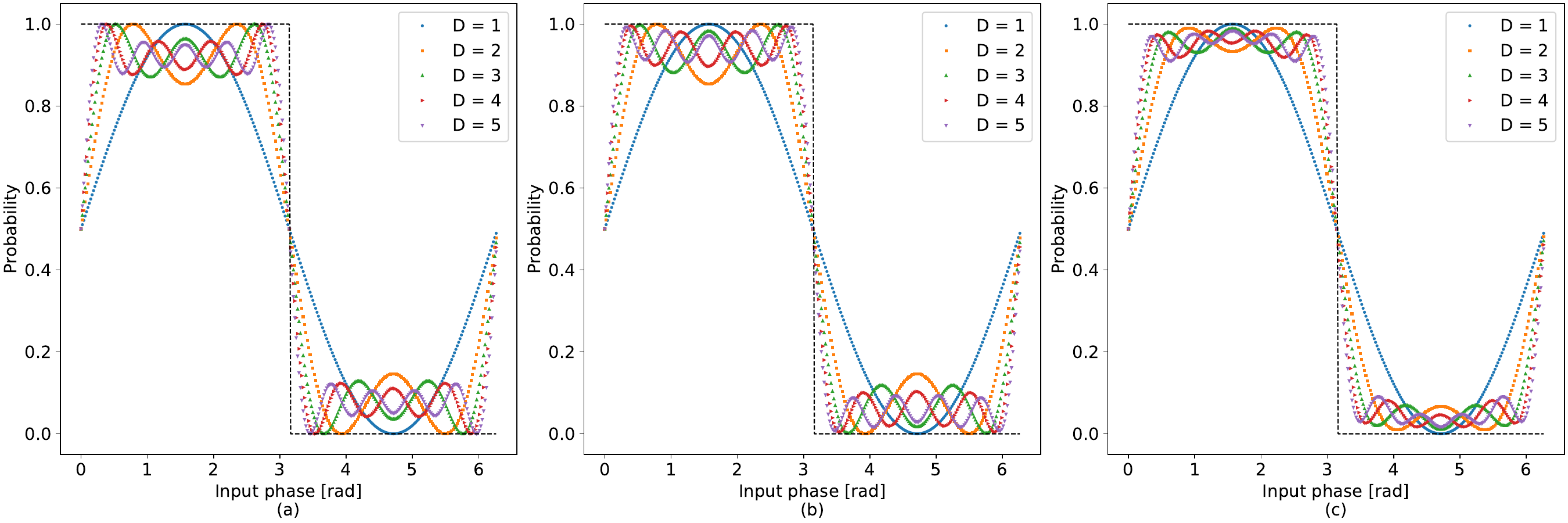}
	\caption{\label{fig:Comparison}Comparison between the square wave $f_j$ and the optical response function $p_{jD}$. Results obtained by direct numerical computation of \cref{eq:StepFunctionFourier} and \cref{eq:Probability}, evaluated for $300$ evenly-spaced phases $\varphi \in [0,2\pi )$, and different truncation orders $D$. The interferometric response is computed exactly, with no statistical sampling method involved. (a) Truncated Fourier series of $f_j$, renormalized to the range $[0,1]$. (b) Interferometer response $p_{jD}$, evaluated for $\beta = 1$ and $\alpha = \pi/2$. (c) Interferometer response $p_{jD}$, evaluated for $\beta = 2$ and $\alpha = \pi/2$.}
\end{figure*}

We complete our analysis by discussing the resource cost of our protocol. In resource counting, the metrological figure of merit is represented by the number $N_p$ of unknown phase shifts $P(\varphi)$, underwent by each photon \citep{art:Giovannetti_Metrology, art:Higgins}. For $k_j = 1$, a single application of $W$ costs $2$ resources. Since $P(k_j\varphi)$ corresponds to $k_j$ compositions of single phase shifts, $W(k_j\varphi, \theta_i)$ requires $2k_j$ resources. After $D$ subsequent applications of $W$, the cost of approximating a binary iteration $f_j$ is $2Dk_j$. The binary search method consists in repeating the above protocol, doubling the value of $k_j$ until condition $\varepsilon = 
\pi/2^n$ is met. The total cost becomes
\begin{equation}
	N_p = 2D\sum_{j=0}^{n-1}k_j = 2D(\pi\varepsilon^{-1}-1) \ , 
\end{equation}
namely $\mathcal{O}(D\pi\varepsilon^{-1})$ resources, with uncertainty $\mathcal{O}(N_p^{-1})$. Our strategy reaches optimality with respect to the Heisenberg bound, with uncertainty obtained by averaging the mean squared error over a single interval of the partition. Notice that protocols that performs below the Heisenberg bound are usually not attainable, since they would require more prior information, making the whole estimation strategy ineffective \citep{art:Giovannetti_SubHeisenberg}.
\begin{algorithm}[H]
\caption{Binary phase estimation\label{alg:Algorithm}}\vspace{4pt}
\textbf{Input:} Phase $\varphi$, uncertainty $\varepsilon$\\[2pt]
\textbf{Output:} Estimation $\widetilde{\varphi}$
\begin{algorithmic}[1]
\State initialize $b \gets (0,\ldots,0)$
\State initialize $j \gets 0$
\While{$\pi / 2^{j} \geq 2\varepsilon$}
	\State measure $p_{jD}(\varphi)$ \Comment{See \cref{eq:Probability}}
	\If{$p_{jD}(\varphi) \geq 1/2$} set $b_j \gets 1$
	\Else{} set $b_j \gets 0$%
	\EndIf%
	\State increment $j \gets j + 1$
\EndWhile
\State get $b \gets \NOT(b)$
\State get $b \gets \REVERSE(b)$
\State convert $\widetilde{m} \gets b_0 2^0 + b_1 2^1 + \ldots + b_{n-1}2^{n-1}$
\State estimate $\widetilde{\varphi}$ \Comment{See \cref{eq:BinaryEstimationOutcome}}
\end{algorithmic}
\end{algorithm}


\section{Conclusions}
We introduced an interferometric setup for the estimation of an unknown phase value. In principle, it can approximate any response function, with Fourier expressivity determined by the depth of the interferometer. In practice, we considered an ad hoc architecture, whose response is a square wave, bypassing the optimization of the interferometer parameters. This choice rephrases the estimation task as a binary search problem, which can estimate the unknown phase at arbitrary uncertainty. Our protocol is compatible with parallel, sequential and multiround strategies. Moreover, it undergoes the Heisenberg scaling in the number of phase transformations, with no prior information required.

\section*{Acknowledgements}
S.R. acknowledges the PRIN MUR Project 2022RATBS4. X.L. thanks National Natural Science Foundation of China under Grant Nos. 62272406, 61932018 and Zhejiang University for funding, and the University of Pavia for hospitality. L.M. acknowledges the U.S. DOE, Office of Science, National Quantum Information Science Research Centers, SQMS Center under Contract No. DE-AC02-07CH11359, and thanks Alexander Sergienko for fruitful discussions.

\section*{Data Availability}
The underlying code that generated the data for this study is openly available in GitHub \citep{rep:BPE}. 

\appendix
\begin{widetext}
\section{Comparison\label{app:Comparison}}
In this section, we investigate the relationship between the square wave $f_j$ and the interferometer output $p_{jD}$. We discuss the role of the parameter $\alpha$ in saturating the response extrema to $0$ and $1$, as shown in \cref{fig:Simulation}. Consider the square wave $f_j$ with period $[0,2\pi/k_j)$ and range $[0,1]$. By direct computation \citep{book:Arfken}, its Fourier series and first-term truncation read
\begin{gather}
	f_{j}(\varphi) = \frac{1}{2} + \sum_{q} \frac{2}{\pi q} \sin\left(q k_j\varphi \right) \ , \label{eq:StepFunctionFourier} \\ 
	f_{j}(\varphi) \simeq \frac{1}{2} + \frac{2}{\pi} \sin\left(k_j\varphi \right) \ , \label{eq:StepFunctionFourierFirst}
\end{gather}
with $q \in 2\mathbb{N} + 1$. Compare this expression against the interferometer response, computed for $D=1$. Consider the set of parameters $\theta_i = \alpha/(\beta i + 1 - \beta)$, with $\alpha, \beta \in \mathbb{R}^+$ and $i \in \{1, \ldots, D\}$. Then
\begin{equation}
\begin{split}
	p_{j1}(\varphi) & = \frac{1}{2} + \frac{1}{8}\Re\left[e^{ik_j\varphi}\left( e^{-i\alpha} - e^{i \alpha} \right)\right] \\
	& = \frac{1}{2} + \frac{1}{2} \sin\alpha \sin(k_j \varphi) \ .
\end{split}
\end{equation}
Independently of $\beta$, we guarantee that $\max p_{j1}(\varphi) = 1$ and $\min p_{j1}(\varphi) = 0$, by choosing $\alpha = \pi/2 \!\! \mod 2\pi$. We numerically evaluate $f_j$ and $p_{jD}$, for different truncation orders $D$, and ensure a representative comparison by renormalizing the truncated Fourier series of $f_j$, so that its range becomes $[0,1]$. The results are reported in \cref{fig:Comparison}, for different values of $\beta$. For $\beta = 1$, $p_{jD}$ better approximates $f_j$. When $\beta = 2$, $p_{jD}$ better saturates its extrema to $0$ and $1$, while mitigating the Gibbs phenomenon.
\end{widetext}

\bibliography{refs.bib}

\begin{thebibliography}{34}%
\makeatletter
\providecommand \@ifxundefined [1]{%
 \@ifx{#1\undefined}
}%
\providecommand \@ifnum [1]{%
 \ifnum #1\expandafter \@firstoftwo
 \else \expandafter \@secondoftwo
 \fi
}%
\providecommand \@ifx [1]{%
 \ifx #1\expandafter \@firstoftwo
 \else \expandafter \@secondoftwo
 \fi
}%
\providecommand \natexlab [1]{#1}%
\providecommand \enquote  [1]{``#1''}%
\providecommand \bibnamefont  [1]{#1}%
\providecommand \bibfnamefont [1]{#1}%
\providecommand \citenamefont [1]{#1}%
\providecommand \href@noop [0]{\@secondoftwo}%
\providecommand \href [0]{\begingroup \@sanitize@url \@href}%
\providecommand \@href[1]{\@@startlink{#1}\@@href}%
\providecommand \@@href[1]{\endgroup#1\@@endlink}%
\providecommand \@sanitize@url [0]{\catcode `\\12\catcode `\$12\catcode
  `\&12\catcode `\#12\catcode `\^12\catcode `\_12\catcode `\%12\relax}%
\providecommand \@@startlink[1]{}%
\providecommand \@@endlink[0]{}%
\providecommand \url  [0]{\begingroup\@sanitize@url \@url }%
\providecommand \@url [1]{\endgroup\@href {#1}{\urlprefix }}%
\providecommand \urlprefix  [0]{URL }%
\providecommand \Eprint [0]{\href }%
\providecommand \doibase [0]{https://doi.org/}%
\providecommand \selectlanguage [0]{\@gobble}%
\providecommand \bibinfo  [0]{\@secondoftwo}%
\providecommand \bibfield  [0]{\@secondoftwo}%
\providecommand \translation [1]{[#1]}%
\providecommand \BibitemOpen [0]{}%
\providecommand \bibitemStop [0]{}%
\providecommand \bibitemNoStop [0]{.\EOS\space}%
\providecommand \EOS [0]{\spacefactor3000\relax}%
\providecommand \BibitemShut  [1]{\csname bibitem#1\endcsname}%
\let\auto@bib@innerbib\@empty
\bibitem [{\citenamefont {Giovannetti}\ \emph {et~al.}(2011)\citenamefont
  {Giovannetti}, \citenamefont {Lloyd},\ and\ \citenamefont
  {Maccone}}]{art:Giovannetti_Advances}%
  \BibitemOpen
  \bibfield  {author} {\bibinfo {author} {\bibfnamefont {V.}~\bibnamefont
  {Giovannetti}}, \bibinfo {author} {\bibfnamefont {S.}~\bibnamefont {Lloyd}},\
  and\ \bibinfo {author} {\bibfnamefont {L.}~\bibnamefont {Maccone}},\
  }\bibfield  {title} {\bibinfo {title} {Advances in quantum metrology},\
  }\href {https://doi.org/10.1038/nphoton.2011.35} {\bibfield  {journal}
  {\bibinfo  {journal} {Nat. Photon.}\ }\textbf {\bibinfo {volume} {5}},\
  \bibinfo {pages} {222–229} (\bibinfo {year} {2011})}\BibitemShut {NoStop}%
\bibitem [{\citenamefont {Paris}(2009)}]{art:Paris}%
  \BibitemOpen
  \bibfield  {author} {\bibinfo {author} {\bibfnamefont {M.~G.~A.}\
  \bibnamefont {Paris}},\ }\bibfield  {title} {\bibinfo {title} {Quantum
  estimation for quantum technology},\ }\href
  {https://doi.org/10.1142/S0219749909004839} {\bibfield  {journal} {\bibinfo
  {journal} {Int. J. Quantum Inf.}\ }\textbf {\bibinfo {volume} {07}},\
  \bibinfo {pages} {125} (\bibinfo {year} {2009})}\BibitemShut {NoStop}%
\bibitem [{\citenamefont {Degen}\ \emph {et~al.}(2017)\citenamefont {Degen},
  \citenamefont {Reinhard},\ and\ \citenamefont {Cappellaro}}]{art:Degen}%
  \BibitemOpen
  \bibfield  {author} {\bibinfo {author} {\bibfnamefont {C.~L.}\ \bibnamefont
  {Degen}}, \bibinfo {author} {\bibfnamefont {F.}~\bibnamefont {Reinhard}},\
  and\ \bibinfo {author} {\bibfnamefont {P.}~\bibnamefont {Cappellaro}},\
  }\bibfield  {title} {\bibinfo {title} {Quantum sensing},\ }\href
  {https://doi.org/10.1103/RevModPhys.89.035002} {\bibfield  {journal}
  {\bibinfo  {journal} {Rev. Mod. Phys.}\ }\textbf {\bibinfo {volume} {89}},\
  \bibinfo {pages} {035002} (\bibinfo {year} {2017})}\BibitemShut {NoStop}%
\bibitem [{\citenamefont {Caves}(1981)}]{art:Caves}%
  \BibitemOpen
  \bibfield  {author} {\bibinfo {author} {\bibfnamefont {C.~M.}\ \bibnamefont
  {Caves}},\ }\bibfield  {title} {\bibinfo {title} {Quantum-mechanical noise in
  an interferometer},\ }\href {https://doi.org/10.1103/PhysRevD.23.1693}
  {\bibfield  {journal} {\bibinfo  {journal} {Phys. Rev. D}\ }\textbf {\bibinfo
  {volume} {23}},\ \bibinfo {pages} {1693} (\bibinfo {year}
  {1981})}\BibitemShut {NoStop}%
\bibitem [{\citenamefont {Kómár}\ \emph {et~al.}(2014)\citenamefont
  {Kómár}, \citenamefont {Kessler}, \citenamefont {Bishof}, \citenamefont
  {Jiang}, \citenamefont {Sørensen}, \citenamefont {Ye},\ and\ \citenamefont
  {Lukin}}]{art:Komar}%
  \BibitemOpen
  \bibfield  {author} {\bibinfo {author} {\bibfnamefont {P.}~\bibnamefont
  {Kómár}}, \bibinfo {author} {\bibfnamefont {E.~M.}\ \bibnamefont
  {Kessler}}, \bibinfo {author} {\bibfnamefont {M.}~\bibnamefont {Bishof}},
  \bibinfo {author} {\bibfnamefont {L.}~\bibnamefont {Jiang}}, \bibinfo
  {author} {\bibfnamefont {A.~S.}\ \bibnamefont {Sørensen}}, \bibinfo {author}
  {\bibfnamefont {J.}~\bibnamefont {Ye}},\ and\ \bibinfo {author}
  {\bibfnamefont {M.~D.}\ \bibnamefont {Lukin}},\ }\bibfield  {title} {\bibinfo
  {title} {A quantum network of clocks},\ }\href
  {https://doi.org/10.1038/nphys3000} {\bibfield  {journal} {\bibinfo
  {journal} {Nat. Phys.}\ }\textbf {\bibinfo {volume} {10}},\ \bibinfo {pages}
  {582–587} (\bibinfo {year} {2014})}\BibitemShut {NoStop}%
\bibitem [{\citenamefont {Bennett}\ and\ \citenamefont
  {Brassard}(2014)}]{art:Bennett}%
  \BibitemOpen
  \bibfield  {author} {\bibinfo {author} {\bibfnamefont {C.~H.}\ \bibnamefont
  {Bennett}}\ and\ \bibinfo {author} {\bibfnamefont {G.}~\bibnamefont
  {Brassard}},\ }\bibfield  {title} {\bibinfo {title} {Quantum cryptography:
  Public key distribution and coin tossing},\ }\href
  {https://doi.org/10.1016/j.tcs.2014.05.025} {\bibfield  {journal} {\bibinfo
  {journal} {Theor. Comput. Sci.}\ }\textbf {\bibinfo {volume} {560}},\
  \bibinfo {pages} {7} (\bibinfo {year} {2014})}\BibitemShut {NoStop}%
\bibitem [{\citenamefont {Tamaki}\ \emph {et~al.}(2012)\citenamefont {Tamaki},
  \citenamefont {Lo}, \citenamefont {Fung},\ and\ \citenamefont
  {Qi}}]{art:Kiyoshi}%
  \BibitemOpen
  \bibfield  {author} {\bibinfo {author} {\bibfnamefont {K.}~\bibnamefont
  {Tamaki}}, \bibinfo {author} {\bibfnamefont {H.-K.}\ \bibnamefont {Lo}},
  \bibinfo {author} {\bibfnamefont {C.-H.~F.}\ \bibnamefont {Fung}},\ and\
  \bibinfo {author} {\bibfnamefont {B.}~\bibnamefont {Qi}},\ }\bibfield
  {title} {\bibinfo {title} {Phase encoding schemes for
  measurement-device-independent quantum key distribution with basis-dependent
  flaw},\ }\href {https://doi.org/10.1103/PhysRevA.85.042307} {\bibfield
  {journal} {\bibinfo  {journal} {Phys. Rev. A}\ }\textbf {\bibinfo {volume}
  {85}},\ \bibinfo {pages} {042307} (\bibinfo {year} {2012})}\BibitemShut
  {NoStop}%
\bibitem [{\citenamefont {Boto}\ \emph {et~al.}(2000)\citenamefont {Boto},
  \citenamefont {Kok}, \citenamefont {Abrams}, \citenamefont {Braunstein},
  \citenamefont {Williams},\ and\ \citenamefont {Dowling}}]{art:Boto}%
  \BibitemOpen
  \bibfield  {author} {\bibinfo {author} {\bibfnamefont {A.~N.}\ \bibnamefont
  {Boto}}, \bibinfo {author} {\bibfnamefont {P.}~\bibnamefont {Kok}}, \bibinfo
  {author} {\bibfnamefont {D.~S.}\ \bibnamefont {Abrams}}, \bibinfo {author}
  {\bibfnamefont {S.~L.}\ \bibnamefont {Braunstein}}, \bibinfo {author}
  {\bibfnamefont {C.~P.}\ \bibnamefont {Williams}},\ and\ \bibinfo {author}
  {\bibfnamefont {J.~P.}\ \bibnamefont {Dowling}},\ }\bibfield  {title}
  {\bibinfo {title} {Quantum interferometric optical lithography: Exploiting
  entanglement to beat the diffraction limit},\ }\href
  {https://doi.org/10.1103/PhysRevLett.85.2733} {\bibfield  {journal} {\bibinfo
   {journal} {Phys. Rev. Lett.}\ }\textbf {\bibinfo {volume} {85}},\ \bibinfo
  {pages} {2733} (\bibinfo {year} {2000})}\BibitemShut {NoStop}%
\bibitem [{\citenamefont {Gessner}\ \emph {et~al.}(2023)\citenamefont
  {Gessner}, \citenamefont {Treps},\ and\ \citenamefont {Fabre}}]{art:Gessner}%
  \BibitemOpen
  \bibfield  {author} {\bibinfo {author} {\bibfnamefont {M.}~\bibnamefont
  {Gessner}}, \bibinfo {author} {\bibfnamefont {N.}~\bibnamefont {Treps}},\
  and\ \bibinfo {author} {\bibfnamefont {C.}~\bibnamefont {Fabre}},\ }\bibfield
   {title} {\bibinfo {title} {Estimation of a parameter encoded in the modal
  structure of a light beam: a quantum theory},\ }\href
  {https://doi.org/10.1364/OPTICA.491368} {\bibfield  {journal} {\bibinfo
  {journal} {Optica}\ }\textbf {\bibinfo {volume} {10}},\ \bibinfo {pages}
  {996} (\bibinfo {year} {2023})}\BibitemShut {NoStop}%
\bibitem [{\citenamefont {Nielsen}\ and\ \citenamefont
  {Chuang}(2010)}]{book:Nielsen}%
  \BibitemOpen
  \bibfield  {author} {\bibinfo {author} {\bibfnamefont {M.~A.}\ \bibnamefont
  {Nielsen}}\ and\ \bibinfo {author} {\bibfnamefont {I.~L.}\ \bibnamefont
  {Chuang}},\ }\href {https://doi.org/10.1017/CBO9780511976667} {\emph
  {\bibinfo {title} {Quantum Computation and Quantum Information: 10th
  Anniversary Edition}}}\ (\bibinfo  {publisher} {Cambridge University Press},\
  \bibinfo {year} {2010})\BibitemShut {NoStop}%
\bibitem [{\citenamefont {Cleve}\ \emph {et~al.}(1998)\citenamefont {Cleve},
  \citenamefont {Ekert}, \citenamefont {Macchiavello},\ and\ \citenamefont
  {Mosca}}]{art:Cleve}%
  \BibitemOpen
  \bibfield  {author} {\bibinfo {author} {\bibfnamefont {R.}~\bibnamefont
  {Cleve}}, \bibinfo {author} {\bibfnamefont {A.}~\bibnamefont {Ekert}},
  \bibinfo {author} {\bibfnamefont {C.}~\bibnamefont {Macchiavello}},\ and\
  \bibinfo {author} {\bibfnamefont {M.}~\bibnamefont {Mosca}},\ }\bibfield
  {title} {\bibinfo {title} {Quantum algorithms revisited},\ }\href
  {https://doi.org/10.1098/rspa.1998.0164} {\bibfield  {journal} {\bibinfo
  {journal} {Proc. R. Soc. Lond. A.}\ }\textbf {\bibinfo {volume} {454}},\
  \bibinfo {pages} {339–354} (\bibinfo {year} {1998})}\BibitemShut {NoStop}%
\bibitem [{\citenamefont {Hassani}\ \emph {et~al.}(2017)\citenamefont
  {Hassani}, \citenamefont {Macchiavello},\ and\ \citenamefont
  {Maccone}}]{art:Hassani}%
  \BibitemOpen
  \bibfield  {author} {\bibinfo {author} {\bibfnamefont {M.}~\bibnamefont
  {Hassani}}, \bibinfo {author} {\bibfnamefont {C.}~\bibnamefont
  {Macchiavello}},\ and\ \bibinfo {author} {\bibfnamefont {L.}~\bibnamefont
  {Maccone}},\ }\bibfield  {title} {\bibinfo {title} {Digital quantum
  estimation},\ }\href {https://doi.org/10.1103/PhysRevLett.119.200502}
  {\bibfield  {journal} {\bibinfo  {journal} {Phys. Rev. Lett.}\ }\textbf
  {\bibinfo {volume} {119}},\ \bibinfo {pages} {200502} (\bibinfo {year}
  {2017})}\BibitemShut {NoStop}%
\bibitem [{\citenamefont {Huerta~Alderete}\ \emph {et~al.}(2022)\citenamefont
  {Huerta~Alderete}, \citenamefont {Gordon}, \citenamefont {Sauvage},
  \citenamefont {Sone}, \citenamefont {Sornborger}, \citenamefont {Coles},\
  and\ \citenamefont {Cerezo}}]{art:Cerezo}%
  \BibitemOpen
  \bibfield  {author} {\bibinfo {author} {\bibfnamefont {C.}~\bibnamefont
  {Huerta~Alderete}}, \bibinfo {author} {\bibfnamefont {M.~H.}\ \bibnamefont
  {Gordon}}, \bibinfo {author} {\bibfnamefont {F.}~\bibnamefont {Sauvage}},
  \bibinfo {author} {\bibfnamefont {A.}~\bibnamefont {Sone}}, \bibinfo {author}
  {\bibfnamefont {A.~T.}\ \bibnamefont {Sornborger}}, \bibinfo {author}
  {\bibfnamefont {P.~J.}\ \bibnamefont {Coles}},\ and\ \bibinfo {author}
  {\bibfnamefont {M.}~\bibnamefont {Cerezo}},\ }\bibfield  {title} {\bibinfo
  {title} {Inference-based quantum sensing},\ }\href
  {https://doi.org/10.1103/PhysRevLett.129.190501} {\bibfield  {journal}
  {\bibinfo  {journal} {Phys. Rev. Lett.}\ }\textbf {\bibinfo {volume} {129}},\
  \bibinfo {pages} {190501} (\bibinfo {year} {2022})}\BibitemShut {NoStop}%
\bibitem [{\citenamefont {Filip}\ \emph {et~al.}(2024)\citenamefont {Filip},
  \citenamefont {Ramo},\ and\ \citenamefont {Fitzpatrick}}]{art:Fitzpatrick}%
  \BibitemOpen
  \bibfield  {author} {\bibinfo {author} {\bibfnamefont {M.-A.}\ \bibnamefont
  {Filip}}, \bibinfo {author} {\bibfnamefont {D.~M.}\ \bibnamefont {Ramo}},\
  and\ \bibinfo {author} {\bibfnamefont {N.}~\bibnamefont {Fitzpatrick}},\
  }\bibfield  {title} {\bibinfo {title} {Variational phase estimation with
  variational fast forwarding},\ }\href
  {https://doi.org/10.22331/q-2024-03-13-1278} {\bibfield  {journal} {\bibinfo
  {journal} {Quantum}\ }\textbf {\bibinfo {volume} {8}},\ \bibinfo {pages}
  {1278} (\bibinfo {year} {2024})}\BibitemShut {NoStop}%
\bibitem [{\citenamefont {Gerry}\ and\ \citenamefont
  {Knight}(2004)}]{book:Gerry}%
  \BibitemOpen
  \bibfield  {author} {\bibinfo {author} {\bibfnamefont {C.}~\bibnamefont
  {Gerry}}\ and\ \bibinfo {author} {\bibfnamefont {P.}~\bibnamefont {Knight}},\
  }\href {https://doi.org/10.1017/CBO9780511791239} {\emph {\bibinfo {title}
  {Introductory Quantum Optics}}}\ (\bibinfo  {publisher} {Cambridge University
  Press},\ \bibinfo {year} {2004})\BibitemShut {NoStop}%
\bibitem [{\citenamefont {Higgins}\ \emph {et~al.}(2007)\citenamefont
  {Higgins}, \citenamefont {Berry}, \citenamefont {Bartlett}, \citenamefont
  {Wiseman},\ and\ \citenamefont {Pryde}}]{art:Higgins}%
  \BibitemOpen
  \bibfield  {author} {\bibinfo {author} {\bibfnamefont {B.~L.}\ \bibnamefont
  {Higgins}}, \bibinfo {author} {\bibfnamefont {D.~W.}\ \bibnamefont {Berry}},
  \bibinfo {author} {\bibfnamefont {S.~D.}\ \bibnamefont {Bartlett}}, \bibinfo
  {author} {\bibfnamefont {H.~M.}\ \bibnamefont {Wiseman}},\ and\ \bibinfo
  {author} {\bibfnamefont {G.~J.}\ \bibnamefont {Pryde}},\ }\bibfield  {title}
  {\bibinfo {title} {Entanglement-free {H}eisenberg-limited phase estimation},\
  }\href {https://doi.org/10.1038/nature06257} {\bibfield  {journal} {\bibinfo
  {journal} {Nature}\ }\textbf {\bibinfo {volume} {450}},\ \bibinfo {pages}
  {393–396} (\bibinfo {year} {2007})}\BibitemShut {NoStop}%
\bibitem [{\citenamefont {Pezz\'e}\ and\ \citenamefont
  {Smerzi}(2008)}]{art:Smerzi}%
  \BibitemOpen
  \bibfield  {author} {\bibinfo {author} {\bibfnamefont {L.}~\bibnamefont
  {Pezz\'e}}\ and\ \bibinfo {author} {\bibfnamefont {A.}~\bibnamefont
  {Smerzi}},\ }\bibfield  {title} {\bibinfo {title} {Mach-{Z}ehnder
  interferometry at the {H}eisenberg limit with coherent and squeezed-vacuum
  light},\ }\href {https://doi.org/10.1103/PhysRevLett.100.073601} {\bibfield
  {journal} {\bibinfo  {journal} {Phys. Rev. Lett.}\ }\textbf {\bibinfo
  {volume} {100}},\ \bibinfo {pages} {073601} (\bibinfo {year}
  {2008})}\BibitemShut {NoStop}%
\bibitem [{\citenamefont {Genoni}\ \emph {et~al.}(2011)\citenamefont {Genoni},
  \citenamefont {Olivares},\ and\ \citenamefont {Paris}}]{art:Olivares}%
  \BibitemOpen
  \bibfield  {author} {\bibinfo {author} {\bibfnamefont {M.~G.}\ \bibnamefont
  {Genoni}}, \bibinfo {author} {\bibfnamefont {S.}~\bibnamefont {Olivares}},\
  and\ \bibinfo {author} {\bibfnamefont {M.~G.~A.}\ \bibnamefont {Paris}},\
  }\bibfield  {title} {\bibinfo {title} {Optical phase estimation in the
  presence of phase diffusion},\ }\href
  {https://doi.org/10.1103/PhysRevLett.106.153603} {\bibfield  {journal}
  {\bibinfo  {journal} {Phys. Rev. Lett.}\ }\textbf {\bibinfo {volume} {106}},\
  \bibinfo {pages} {153603} (\bibinfo {year} {2011})}\BibitemShut {NoStop}%
\bibitem [{\citenamefont {Schwarze}\ \emph {et~al.}(2023)\citenamefont
  {Schwarze}, \citenamefont {Simon},\ and\ \citenamefont
  {Sergienko}}]{art:Sergienko}%
  \BibitemOpen
  \bibfield  {author} {\bibinfo {author} {\bibfnamefont {C.~R.}\ \bibnamefont
  {Schwarze}}, \bibinfo {author} {\bibfnamefont {D.~S.}\ \bibnamefont
  {Simon}},\ and\ \bibinfo {author} {\bibfnamefont {A.~V.}\ \bibnamefont
  {Sergienko}},\ }\bibfield  {title} {\bibinfo {title} {Enhanced-sensitivity
  interferometry with phase-sensitive unbiased multiports},\ }\href
  {https://doi.org/10.1103/PhysRevA.107.052615} {\bibfield  {journal} {\bibinfo
   {journal} {Phys. Rev. A}\ }\textbf {\bibinfo {volume} {107}},\ \bibinfo
  {pages} {052615} (\bibinfo {year} {2023})}\BibitemShut {NoStop}%
\bibitem [{\citenamefont {Sinanan-Singh}\ \emph {et~al.}(2024)\citenamefont
  {Sinanan-Singh}, \citenamefont {Mintzer}, \citenamefont {Chuang},\ and\
  \citenamefont {Liu}}]{art:Chuang}%
  \BibitemOpen
  \bibfield  {author} {\bibinfo {author} {\bibfnamefont {J.}~\bibnamefont
  {Sinanan-Singh}}, \bibinfo {author} {\bibfnamefont {G.~L.}\ \bibnamefont
  {Mintzer}}, \bibinfo {author} {\bibfnamefont {I.~L.}\ \bibnamefont
  {Chuang}},\ and\ \bibinfo {author} {\bibfnamefont {Y.}~\bibnamefont {Liu}},\
  }\bibfield  {title} {\bibinfo {title} {Single-shot quantum signal processing
  interferometry},\ }\href {https://doi.org/10.22331/q-2024-07-30-1427}
  {\bibfield  {journal} {\bibinfo  {journal} {Quantum}\ }\textbf {\bibinfo
  {volume} {8}},\ \bibinfo {pages} {1427} (\bibinfo {year} {2024})}\BibitemShut
  {NoStop}%
\bibitem [{\citenamefont {Wiebe}\ and\ \citenamefont
  {Granade}(2016)}]{art:Wiebe}%
  \BibitemOpen
  \bibfield  {author} {\bibinfo {author} {\bibfnamefont {N.}~\bibnamefont
  {Wiebe}}\ and\ \bibinfo {author} {\bibfnamefont {C.}~\bibnamefont
  {Granade}},\ }\bibfield  {title} {\bibinfo {title} {Efficient bayesian phase
  estimation},\ }\href {https://doi.org/10.1103/PhysRevLett.117.010503}
  {\bibfield  {journal} {\bibinfo  {journal} {Phys. Rev. Lett.}\ }\textbf
  {\bibinfo {volume} {117}},\ \bibinfo {pages} {010503} (\bibinfo {year}
  {2016})}\BibitemShut {NoStop}%
\bibitem [{\citenamefont {Busch}\ \emph {et~al.}(2007)\citenamefont {Busch},
  \citenamefont {Heinonen},\ and\ \citenamefont {Lahti}}]{art:Busch}%
  \BibitemOpen
  \bibfield  {author} {\bibinfo {author} {\bibfnamefont {P.}~\bibnamefont
  {Busch}}, \bibinfo {author} {\bibfnamefont {T.}~\bibnamefont {Heinonen}},\
  and\ \bibinfo {author} {\bibfnamefont {P.}~\bibnamefont {Lahti}},\ }\bibfield
   {title} {\bibinfo {title} {Heisenberg's uncertainty principle},\ }\href
  {https://doi.org/10.1016/j.physrep.2007.05.006} {\bibfield  {journal}
  {\bibinfo  {journal} {Phys. Rep.}\ }\textbf {\bibinfo {volume} {452}},\
  \bibinfo {pages} {155} (\bibinfo {year} {2007})}\BibitemShut {NoStop}%
\bibitem [{\citenamefont {Giovannetti}\ \emph {et~al.}(2004)\citenamefont
  {Giovannetti}, \citenamefont {Lloyd},\ and\ \citenamefont
  {Maccone}}]{art:Giovannetti_Measurements}%
  \BibitemOpen
  \bibfield  {author} {\bibinfo {author} {\bibfnamefont {V.}~\bibnamefont
  {Giovannetti}}, \bibinfo {author} {\bibfnamefont {S.}~\bibnamefont {Lloyd}},\
  and\ \bibinfo {author} {\bibfnamefont {L.}~\bibnamefont {Maccone}},\
  }\bibfield  {title} {\bibinfo {title} {Quantum-enhanced measurements: Beating
  the standard quantum limit},\ }\href
  {https://doi.org/10.1126/science.1104149} {\bibfield  {journal} {\bibinfo
  {journal} {Science}\ }\textbf {\bibinfo {volume} {306}},\ \bibinfo {pages}
  {1330} (\bibinfo {year} {2004})}\BibitemShut {NoStop}%
\bibitem [{\citenamefont {G\'orecki}\ \emph {et~al.}(2020)\citenamefont
  {G\'orecki}, \citenamefont {Demkowicz-Dobrza\ifmmode~\acute{n}\else
  \'{n}\fi{}ski}, \citenamefont {Wiseman},\ and\ \citenamefont
  {Berry}}]{art:Gorecki}%
  \BibitemOpen
  \bibfield  {author} {\bibinfo {author} {\bibfnamefont {W.}~\bibnamefont
  {G\'orecki}}, \bibinfo {author} {\bibfnamefont {R.}~\bibnamefont
  {Demkowicz-Dobrza\ifmmode~\acute{n}\else \'{n}\fi{}ski}}, \bibinfo {author}
  {\bibfnamefont {H.~M.}\ \bibnamefont {Wiseman}},\ and\ \bibinfo {author}
  {\bibfnamefont {D.~W.}\ \bibnamefont {Berry}},\ }\bibfield  {title} {\bibinfo
  {title} {$\ensuremath{\pi}$-{C}orrected {H}eisenberg limit},\ }\href
  {https://doi.org/10.1103/PhysRevLett.124.030501} {\bibfield  {journal}
  {\bibinfo  {journal} {Phys. Rev. Lett.}\ }\textbf {\bibinfo {volume} {124}},\
  \bibinfo {pages} {030501} (\bibinfo {year} {2020})}\BibitemShut {NoStop}%
\bibitem [{\citenamefont {Belliardo}\ and\ \citenamefont
  {Giovannetti}(2020)}]{art:Belliardo}%
  \BibitemOpen
  \bibfield  {author} {\bibinfo {author} {\bibfnamefont {F.}~\bibnamefont
  {Belliardo}}\ and\ \bibinfo {author} {\bibfnamefont {V.}~\bibnamefont
  {Giovannetti}},\ }\bibfield  {title} {\bibinfo {title} {Achieving heisenberg
  scaling with maximally entangled states: An analytic upper bound for the
  attainable root-mean-square error},\ }\href
  {https://doi.org/10.1103/PhysRevA.102.042613} {\bibfield  {journal} {\bibinfo
   {journal} {Phys. Rev. A}\ }\textbf {\bibinfo {volume} {102}},\ \bibinfo
  {pages} {042613} (\bibinfo {year} {2020})}\BibitemShut {NoStop}%
\bibitem [{\citenamefont {Maccone}\ and\ \citenamefont
  {De~Cillis}(2009)}]{art:DeCillis}%
  \BibitemOpen
  \bibfield  {author} {\bibinfo {author} {\bibfnamefont {L.}~\bibnamefont
  {Maccone}}\ and\ \bibinfo {author} {\bibfnamefont {G.}~\bibnamefont
  {De~Cillis}},\ }\bibfield  {title} {\bibinfo {title} {Robust strategies for
  lossy quantum interferometry},\ }\href
  {https://doi.org/10.1103/PhysRevA.79.023812} {\bibfield  {journal} {\bibinfo
  {journal} {Phys. Rev. A}\ }\textbf {\bibinfo {volume} {79}},\ \bibinfo
  {pages} {023812} (\bibinfo {year} {2009})}\BibitemShut {NoStop}%
\bibitem [{\citenamefont {Lee}\ \emph {et~al.}(2002)\citenamefont {Lee},
  \citenamefont {Kok},\ and\ \citenamefont {Dowling}}]{art:Lee}%
  \BibitemOpen
  \bibfield  {author} {\bibinfo {author} {\bibfnamefont {H.}~\bibnamefont
  {Lee}}, \bibinfo {author} {\bibfnamefont {P.}~\bibnamefont {Kok}},\ and\
  \bibinfo {author} {\bibfnamefont {J.~P.}\ \bibnamefont {Dowling}},\
  }\bibfield  {title} {\bibinfo {title} {A quantum {R}osetta stone for
  interferometry},\ }\href {https://doi.org/10.1080/0950034021000011536}
  {\bibfield  {journal} {\bibinfo  {journal} {J. Mod. Opt.}\ }\textbf {\bibinfo
  {volume} {49}},\ \bibinfo {pages} {2325–2338} (\bibinfo {year}
  {2002})}\BibitemShut {NoStop}%
\bibitem [{\citenamefont {Stone}(1932)}]{art:Stone}%
  \BibitemOpen
  \bibfield  {author} {\bibinfo {author} {\bibfnamefont {M.~H.}\ \bibnamefont
  {Stone}},\ }\bibfield  {title} {\bibinfo {title} {On one-parameter unitary
  groups in hilbert space},\ }\href {https://doi.org/10.2307/1968538}
  {\bibfield  {journal} {\bibinfo  {journal} {Ann. Math.}\ }\textbf {\bibinfo
  {volume} {33}},\ \bibinfo {pages} {643} (\bibinfo {year} {1932})}\BibitemShut
  {NoStop}%
\bibitem [{\citenamefont {Schuld}\ \emph {et~al.}(2021)\citenamefont {Schuld},
  \citenamefont {Sweke},\ and\ \citenamefont {Meyer}}]{art:Schuld}%
  \BibitemOpen
  \bibfield  {author} {\bibinfo {author} {\bibfnamefont {M.}~\bibnamefont
  {Schuld}}, \bibinfo {author} {\bibfnamefont {R.}~\bibnamefont {Sweke}},\ and\
  \bibinfo {author} {\bibfnamefont {J.~J.}\ \bibnamefont {Meyer}},\ }\bibfield
  {title} {\bibinfo {title} {Effect of data encoding on the expressive power of
  variational quantum-machine-learning models},\ }\href
  {https://doi.org/10.1103/PhysRevA.103.032430} {\bibfield  {journal} {\bibinfo
   {journal} {Phys. Rev. A}\ }\textbf {\bibinfo {volume} {103}},\ \bibinfo
  {pages} {032430} (\bibinfo {year} {2021})}\BibitemShut {NoStop}%
\bibitem [{\citenamefont {Schuld}\ and\ \citenamefont
  {Petruccione}(2021)}]{book:Petruccione}%
  \BibitemOpen
  \bibfield  {author} {\bibinfo {author} {\bibfnamefont {M.}~\bibnamefont
  {Schuld}}\ and\ \bibinfo {author} {\bibfnamefont {F.}~\bibnamefont
  {Petruccione}},\ }\href {https://doi.org/10.1007/978-3-030-83098-4} {\emph
  {\bibinfo {title} {Machine Learning with Quantum Computers}}}\ (\bibinfo
  {publisher} {Springer},\ \bibinfo {year} {2021})\BibitemShut {NoStop}%
\bibitem [{\citenamefont {Giovannetti}\ \emph {et~al.}(2006)\citenamefont
  {Giovannetti}, \citenamefont {Lloyd},\ and\ \citenamefont
  {Maccone}}]{art:Giovannetti_Metrology}%
  \BibitemOpen
  \bibfield  {author} {\bibinfo {author} {\bibfnamefont {V.}~\bibnamefont
  {Giovannetti}}, \bibinfo {author} {\bibfnamefont {S.}~\bibnamefont {Lloyd}},\
  and\ \bibinfo {author} {\bibfnamefont {L.}~\bibnamefont {Maccone}},\
  }\bibfield  {title} {\bibinfo {title} {Quantum metrology},\ }\href
  {https://doi.org/10.1103/PhysRevLett.96.010401} {\bibfield  {journal}
  {\bibinfo  {journal} {Phys. Rev. Lett.}\ }\textbf {\bibinfo {volume} {96}},\
  \bibinfo {pages} {010401} (\bibinfo {year} {2006})}\BibitemShut {NoStop}%
\bibitem [{\citenamefont {Giovannetti}\ and\ \citenamefont
  {Maccone}(2012)}]{art:Giovannetti_SubHeisenberg}%
  \BibitemOpen
  \bibfield  {author} {\bibinfo {author} {\bibfnamefont {V.}~\bibnamefont
  {Giovannetti}}\ and\ \bibinfo {author} {\bibfnamefont {L.}~\bibnamefont
  {Maccone}},\ }\bibfield  {title} {\bibinfo {title} {Sub-{H}eisenberg
  estimation strategies are ineffective},\ }\href
  {https://doi.org/10.1103/PhysRevLett.108.210404} {\bibfield  {journal}
  {\bibinfo  {journal} {Phys. Rev. Lett.}\ }\textbf {\bibinfo {volume} {108}},\
  \bibinfo {pages} {210404} (\bibinfo {year} {2012})}\BibitemShut {NoStop}%
\bibitem [{rep()}]{rep:BPE}%
  \BibitemOpen
  \href@noop {} {}\bibinfo {note}
  {\url{https://github.com/simoneroncallo/binary-phase-estimation}}\BibitemShut
  {NoStop}%
\bibitem [{\citenamefont {Arfken}\ \emph {et~al.}(2012)\citenamefont {Arfken},
  \citenamefont {Arfken}, \citenamefont {Weber},\ and\ \citenamefont
  {Harris}}]{book:Arfken}%
  \BibitemOpen
  \bibfield  {author} {\bibinfo {author} {\bibfnamefont {G.}~\bibnamefont
  {Arfken}}, \bibinfo {author} {\bibfnamefont {G.}~\bibnamefont {Arfken}},
  \bibinfo {author} {\bibfnamefont {H.}~\bibnamefont {Weber}},\ and\ \bibinfo
  {author} {\bibfnamefont {F.}~\bibnamefont {Harris}},\ }\href
  {https://doi.org/10.1016/C2009-0-30629-7} {\emph {\bibinfo {title}
  {Mathematical Methods for Physicists: A Comprehensive Guide}}}\ (\bibinfo
  {publisher} {Elsevier},\ \bibinfo {year} {2012})\BibitemShut {NoStop}%
\end{thebibliography}%
\end{document}